\documentclass[12pt]{article}

\usepackage{graphicx,amssymb,amsmath,dsfont,txfonts} 
\usepackage{bbm} 
\usepackage{mathtools} 

\usepackage[a4paper]{geometry} 
\def\dfrac#1#2{\frac{\displaystyle\strut #1}{\displaystyle\strut #2}}
\DeclareMathOperator{\tr}{tr}
\def\bra#1{\mathinner{\langle{#1}|}}
\def\ket#1{\mathinner{|{#1}\rangle}}
\def\braket#1{\mathinner{\langle{#1}\rangle}}

\begin{document}

\title{\LARGE\bf 
Indefinite causal order for quantum metrology \\ with quantum thermal noise}

\author{Fran\c{c}ois {\sc Chapeau-Blondeau}, \\
    Laboratoire Angevin de Recherche en Ing\'enierie des Syst\`emes (LARIS), \\
    Universit\'e d'Angers,
    62 avenue Notre Dame du Lac, 49000 Angers, France.
}

\date{\today}

\maketitle

\parindent=8mm \parskip=0ex

\begin{abstract}
A switched quantum channel with indefinite causal order is studied for the fundamental 
metrological task of phase estimation on a qubit unitary operator affected by quantum thermal 
noise. Specific capabilities are reported in the switched channel with indefinite order,
not accessible with conventional estimation approaches with definite order.
Phase estimation can be performed by measuring the control qubit alone, although
it does not actively interact with the unitary process -- only the probe qubit doing so.
Also, phase estimation becomes possible with a fully depolarized input probe or with an
input probe aligned with the rotation axis of the unitary, while this is never possible with 
conventional approaches. The present study extends to thermal noise, investigations previously 
carried out with the more symmetric and isotropic qubit depolarizing noise, and it contributes 
to the timely exploration of properties of quantum channels with indefinite causal order relevant 
to quantum signal and information processing.
\end{abstract}

\maketitle

\section{Introduction}

{\let\thefootnote\relax\footnote{{Preprint of a paper published by {\em Physics Letters A},
vol.~447, 128300, pp.~1--10 (2022). \\
https://doi.org/10.1016/j.physleta.2022.128300 \hfill 
\;\;\;https://www.sciencedirect.com/science/article/abs/pii/S0375960122003826}}}
Quantum channels arranged in indefinite causal order represent novel architectures for combining
quantum processes, which exhibit specific properties useful to quantum signal and information 
processing, and not accessible with conventional associations with definite causal order.
The principle of quantum channels with indefinite causal order has been analyzed in 
\cite{Oreshkov12,Chiribella13}, and their physical implementation is discussed for instance in 
\cite{Chiribella13,Procopio15,Rubino17,Goswami18,Wei19,Guo20}.
Quantum channels with indefinite causal order have been shown profitable in various tasks of
quantum information processing, such as communication over noisy quantum channels 
\cite{Ebler18,Procopio19,Procopio20,Loizeau20}, or quantum channel discrimination 
\cite{Chiribella12,Koudia19}. For quantum metrology, which will be the principal concern of the
present study, channels with indefinite order have been studied in 
\cite{Frey19,Mukhopadhyay18,Zhao20,Chapeau21}, to estimate the level of a qudit depolarizing noise 
in \cite{Frey19} or the temperature of a qubit thermal noise in \cite{Mukhopadhyay18}, or to extract
information about average displacements in a quantum system with continuous variables in 
\cite{Zhao20}. The study of \cite{Chapeau21} addresses the reference task of quantum metrology 
consisting in quantum phase estimation in the presence of noise. Phase estimation is a fundamental 
task of quantum metrology, useful to many applications related to high-sensitivity and high-precision 
physical measurements \cite{Giovannetti06,Giovannetti11,DAriano98,vanDam07,Chapeau15,Degen17}. 
For qubit phase estimation in the presence of a qubit depolarizing noise,
Ref.~\cite{Chapeau21} demonstrates various capabilities afforded by a switched quantum
channel with indefinite causal order, and that are not accessible with standard estimation
approaches with definite order.
In the present study, we will address the same type of switched quantum channels
with indefinite causal order involved in a task of phase estimation, as in \cite{Chapeau21}.
While \cite{Chapeau21} considers phase estimation in the presence of a unital qubit depolarizing 
noise showing high symmetry and isotropy, we will extend the analysis here to a less regular
nonunital quantum noise under the form of a qubit thermal noise, which represents also a noise of 
high practical relevance \cite{Oliveira20,Khatri20,Falaye17,Tham16,Wang14}.
The present report provides the first analysis of a switched quantum channel with indefinite 
causal order for quantum phase estimation in the presence of thermal noise. It contributes to the 
inventory and appreciation of specific properties of quantum channels with indefinite causal order 
bearing relevance to quantum signal and information processing.

\section{Controlled switch of two quantum channels} \label{switch_sec}

\begin{figure}[htb]
\centerline{\includegraphics[width=75mm]{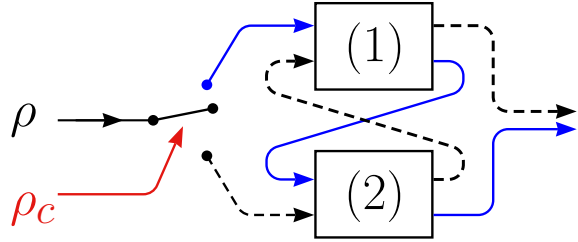}} 
\caption[what appears in lof LL p177]
{According to the state $\rho_c$ of a control qubit, the two quantum channels (1) and (2) can be 
cascaded in the causal order (1)--(2) (solid path) or (2)--(1) (dashed path) to affect the probe 
signal in state $\rho$.
}
\label{figSwi1}
\end{figure}

We consider here, as in Refs.~\cite{Chiribella13,Ebler18,Chapeau21}, two quantum channels (1) and 
(2), respectively characterized by the Kraus operators $\bigl\{\mathsf{K}_k^{(1)} \bigr\}$ and
$\bigl\{\mathsf{K}_k^{(2)} \bigr\}$, and which are switched between the two causal orders (1)--(2) 
or (2)--(1) according to the state $\ket{0_c}$ or $\ket{1_c}$ or a control qubit,
as represented in Fig.~\ref{figSwi1}

A switched quantum channel results, which is characterized \cite{Chiribella13,Ebler18,Chapeau21} 
by the Kraus operators 
\begin{equation}
\mathsf{K}_{jk} = \mathsf{K}_j^{(2)} \mathsf{K}_k^{(1)} \otimes \ket{0_c}\bra{0_c} + 
                  \mathsf{K}_k^{(1)} \mathsf{K}_j^{(2)} \otimes \ket{1_c}\bra{1_c} \;.
\label{Wjk}
\end{equation}
With a control qubit placed in the general state represented by the density operator $\rho_c$,
the switched quantum channel of Fig.~\ref{figSwi1} realizes the bipartite transformation
\begin{equation}
\mathcal{S}(\rho \otimes \rho_c) = \sum_{j} \sum_k \mathsf{K}_{jk} (\rho \otimes \rho_c) 
\mathsf{K}_{jk}^\dagger \;.
\label{Sgen1}
\end{equation}
An interesting possibility is to place the control qubit in a coherent superposition of its
two basis states, reading $\ket{\psi_c}=\sqrt{p_c}\ket{0_c}+\sqrt{1-p_c}\ket{1_c}$, with 
$p_c \in [0, 1]$.
In this way, the switched channel of Fig.~\ref{figSwi1} realizes a coherent superposition
of the two alternative causal orders (1)--(2) and (2)--(1), implementing a switched quantum channel 
with indefinite causal order. In the general case of a control qubit with density operator $\rho_c$, 
pure or mixed, the transformation of Eq.~(\ref{Sgen1}) can be developed \cite{Chapeau21} into
\begin{eqnarray}
\nonumber
\mathcal{S}(\rho \otimes \rho_c) &=& 
\mathcal{S}_{00}(\rho) \otimes \braket{0_c|\rho_c|0_c}\ket{0_c}\bra{0_c} +
\mathcal{S}_{01}(\rho) \otimes \braket{0_c|\rho_c|1_c}\ket{0_c}\bra{1_c} \\
\label{Sgen2} 
\mbox{} &+& \mathcal{S}_{01}^\dagger(\rho) \otimes \braket{1_c|\rho_c|0_c}\ket{1_c}\bra{0_c} +
\mathcal{S}_{11}(\rho) \otimes \braket{1_c|\rho_c|1_c}\ket{1_c}\bra{1_c} \;.
\end{eqnarray}
In Eq.~(\ref{Sgen2}), the superoperators $\mathcal{S}_{00}(\rho)$ and $\mathcal{S}_{11}(\rho)$ 
respectively describe transmission by the standard cascades (1)--(2) and (2)--(1), defined by the 
two sets of Kraus operators $\bigl\{\mathsf{K}_j^{(2)} \mathsf{K}_k^{(1)} \bigr\}$ and
$\bigl\{ \mathsf{K}_k^{(1)} \mathsf{K}_j^{(2)} \bigr\}$. 
With the pure state $\rho_c=\ket{\psi_c}\bra{\psi_c}$, at $p_c=0$ or $p_c=1$ there is no genuine 
superposition of causal orders, and the operation of the switched channel in Eq.~(\ref{Sgen2})
reduces to these standard cascades (1)--(2) or (2)--(1),
via $\mathcal{S}_{00}(\rho)$ or $\mathcal{S}_{11}(\rho)$. 
By contrast in Eq.~(\ref{Sgen2}), the superoperator 
\begin{equation}
\mathcal{S}_{01}(\rho) = \sum_j \sum_k \mathsf{K}_j^{(2)} \mathsf{K}_k^{(1)} \rho 
\mathsf{K}_j^{(2)\dagger} \mathsf{K}_k^{(1)\dagger} \;,
\label{S01}
\end{equation}
specifically conveys the effect of the superposition of causal orders, especially acting
with $\rho_c=\ket{\psi_c}\bra{\psi_c}$ when $p_c \not = 0$ and $p_c \not = 1$.

We will now specifically investigate the situation where the two superposed channels (1) and (2) 
in Fig.~\ref{figSwi1} are formed by a unitary qubit channel affected by quantum noise,
to be involved in the fundamental metrological task of phase estimation on the unitary.
For such a scenario of phase estimation in a switched quantum channel with indefinite causal order,
\cite{Chapeau21} studied the case of the unital isotropic depolarizing noise, while here we
will investigate the case of a nonunital and less symmetric thermal noise, having also great
practical relevance for the qubit \cite{Oliveira20,Khatri20,Falaye17,Tham16,Wang14}.

\section{A switched qubit unitary channel with thermal noise}

\begin{figure}[htb]
\centerline{\includegraphics[width=75mm]{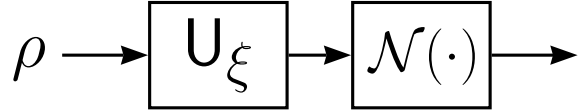}} 
\caption[what appears in lof LL p177]
{Quantum channel formed by a qubit unitary operator $\mathsf{U}_\xi$ affected by a qubit noise 
$\mathcal{N}(\cdot)$, and as a whole providing a realization for channel (1) or (2) involved 
in Fig.~\ref{figSwi1}.
}
\label{fig_blocUN}
\end{figure}

We will examine a noisy quantum channel consisting in a qubit unitary operator $\mathsf{U}_\xi$
affected by a qubit noise $\mathcal{N}(\cdot)$ as represented in Fig.~\ref{fig_blocUN}.

The quantum channel of Fig.~\ref{fig_blocUN} acts on a probe qubit with input density operator
\begin{equation}
\rho = \left[ \begin{array}{cc}
\rho_{00} & \rho_{01}  \\
\rho_{01}^* & 1-\rho_{00}
\end{array} \right] = 
\frac{1}{2}\bigl( \mathrm{I}_2 + \vec{r}\cdot \vec{\sigma} \bigr) \;,
\label{ro_in1}
\end{equation}
with in particular $\rho_{00} \in [0, 1]$. In Eq.~(\ref{ro_in1}), $\mathrm{I}_2$ is the identity 
operator on the two-dimensional qubit space $\mathcal{H}_2$, and $\vec{\sigma}$ is the formal 
vector of the three Pauli operators $[\sigma_x, \sigma_y, \sigma_z]=\vec{\sigma}$. The Bloch vector 
$\vec{r} \in \mathbbm{R}^3$  is with norm $\| \vec{r}\, \|=1$ for a pure state, and 
$\| \vec{r}\, \|<1$ for a mixed state.

The qubit unitary operator $\mathsf{U}_\xi$ of Fig.~\ref{fig_blocUN} receives \cite{Nielsen00} the 
general parameterization  
\begin{equation}
\mathsf{U}_\xi=\exp\Bigl(-i\dfrac{\xi}{2} \vec{n} \cdot \vec{\sigma} \Bigr) \;,
\label{Uxi}
\end{equation}
with $\vec{n}=[n_x, n_y, n_z]^\top$ a unit vector of $\mathbbm{R}^3$, and $\xi$ a phase angle in 
$[0, 2\pi )$ which is often a parameter of metrological interest. The unitary implements the 
transformation $\rho \mapsto \mathsf{U}_\xi \rho \mathsf{U}_\xi^\dagger$, equivalent to
transforming the Bloch vector of the qubit by $\vec{r} \mapsto U_\xi \vec{r}$ with $U_\xi$ 
(in italic) the $3\times 3$ rotation matrix of $\mathbbm{R}^3$ of angle $\xi$ around the axis 
$\vec{n}$.

The quantum noise $\mathcal{N}(\cdot)$ of Fig.~\ref{fig_blocUN} is taken here as a generalized 
amplitude damping noise or quantum thermal noise, defined \cite{Nielsen00} by the four Kraus 
operators 
\begin{eqnarray}
\label{4.tgad1}
\Lambda_1 &=& \sqrt{p}
\left[ \begin{array}{cc}
1 & 0  \\
0 & \sqrt{1-\gamma}
\end{array} \right] \;, \\
\label{4.tgad2}
\Lambda_2 &=& \sqrt{p}
\left[ \begin{array}{cc}
0 & \sqrt{\gamma}  \\
0 & 0
\end{array} \right] \;, \\
\label{4.tgad3}
\Lambda_3 &=& \sqrt{1-p}
\left[ \begin{array}{cc}
\sqrt{1-\gamma} & 0  \\
0               & 1  
\end{array} \right] \;, \\
\label{4.tgad4}
\Lambda_4 &=& \sqrt{1-p}
\left[ \begin{array}{cc}
0             & 0  \\
\sqrt{\gamma} & 0
\end{array} \right] \;.
\end{eqnarray}
On the qubit density operator $\rho$ of Eq.~(\ref{ro_in1}), the noise implements the (nonunitary) 
transformation $\rho \mapsto \mathcal{N}(\rho)=\sum_{j=1}^4 \Lambda_j \rho \Lambda_j^\dagger$ 
reading
\begin{equation}
\mathcal{N}(\rho)=
\left[ \begin{array}{lr} 
(1-\gamma)\rho_{00}+p\gamma & \sqrt{1-\gamma}\, \rho_{01} \\
\sqrt{1-\gamma}\, \rho_{01}^* & (1-\gamma)(1-\rho_{00})+(1-p)\gamma
\end{array} \right] \;,
\label{4.A_NGAD}
\end{equation}
equivalent to transforming the Bloch vector of the qubit by 
\begin{equation}
\vec{r} \longmapsto A \vec{r}+ \vec{c}=
\left[ \begin{array}{ccc} 
\sqrt{1-\gamma} & 0               & 0 \\
0               & \sqrt{1-\gamma} & 0 \\
0               & 0               & 1-\gamma 
\end{array} \right]\vec{r} +
\left[ \begin{array}{c} 
0 \\
0 \\
(2p-1)\gamma 
\end{array} \right] \;.
\label{4.A_GAD}
\end{equation}

Such noise offers a model \cite{Nielsen00,Falaye17} to describe the coupling of the qubit with an 
uncontrolled environment consisting in a thermal bath at temperature $T$. The damping factor 
$\gamma \in [0, 1]$ can be related to the interaction time $t$ of the qubit with the bath as 
$\gamma = 1-e^{-t/\tau_1}$, with $\tau_1$ a relaxation time for the interaction. At 
long time $t \gg \tau_1$, one has $\gamma \rightarrow 1$ and a qubit relaxing to the 
equilibrium or thermalized mixed state $\rho_\infty= p \ket{0}\bra{0}+ (1-p) \ket{1}\bra{1}$ having
Bloch vector $\vec{r}_\infty =[0, 0, 2p-1]^\top$. At equilibrium, the probability is $p$ for
measuring the qubit in the ground state $\ket{0}$ and $1-p$ for measuring it in the excited state 
$\ket{1}$. With the energies $E_0$ and $E_1 > E_0$ respectively for the states $\ket{0}$ and 
$\ket{1}$, the equilibrium probability $p$ is given by the Boltzmann distribution as
\begin{equation}
p=\dfrac{1}{1+\exp[-(E_1-E_0)/(k_BT)]} \;.
\label{4.pT}
\end{equation}
By Eq.~(\ref{4.pT}), the probability $p$ characterizing the quantum thermal noise of 
Eqs.~(\ref{4.tgad1})--(\ref{4.A_GAD}), is determined by the temperature $T$ of the bath. From 
Eq.~(\ref{4.pT}), the probability $p$ decreases as the temperature $T$ increases. A temperature $T=0$ 
gives a probability $p=1$ for the ground state $\ket{0}$, while at $T\rightarrow \infty$ the 
ground state $\ket{0}$ and excited state $\ket{1}$ become equiprobable with $p=1/2$. Therefore, 
from Eq.~(\ref{4.pT}), when the temperature $T$ monotonically increases from $0$ to $\infty$, the 
probability $p$ monotonically decreases from $1$ to $1/2$. 
For the sequel, as in \cite{Gillard18b}, it will be convenient to refer to an effective or reduced 
noise temperature $T_p=2(1-p)$, which, according to Eq.~(\ref{4.pT}), is a monotonically increasing 
function of the physical temperature $T$, for any value of the energy gap $E_1-E_0 >0$. When 
$T=0$ then $p=1$ and $T_p=0$, while when $T=\infty$ then $p=1/2$ and $T_p=1$. 
So a physical temperature $T$ increasing from $0$ to $\infty$ is monotonically mapped into an 
effective temperature $T_p$ increasing from $0$ to $1$. This provides a convenient finite 
range of $T_p \in [0, 1]$ to convey the impact of the physical temperature 
$T\in[0, \infty [\,$, and also releases the quantitative analysis from the unimportant specific 
value of the energy gap $E_1-E_0 >0$.

The quantum channel resulting in Fig.~\ref{fig_blocUN} is therefore defined by the four Kraus 
operators $\mathsf{K}_j=\Lambda_j \mathsf{U}_\xi$, for $j=1$ to $4$. It implements the 
transformation 
$\rho \mapsto \sum_{j=1}^4 \Lambda_j \mathsf{U}_\xi \rho \mathsf{U}_\xi^\dagger \Lambda_j^\dagger$, 
equivalent to transforming the Bloch vector of the qubit by 
$\vec{r} \mapsto AU_\xi \vec{r}+\vec{c}$.
Two such identical noisy unitary channels as in Fig.~\ref{fig_blocUN} are employed to realize 
channels (1) and (2) in the switched quantum channel of Fig.~\ref{figSwi1}. Two identical channels 
(1) and (2) in Fig.~\ref{figSwi1} give $\mathcal{S}_{00}(\rho)=\mathcal{S}_{11}(\rho)$ and 
$\mathcal{S}_{01}^\dagger(\rho)=\mathcal{S}_{01}(\rho)$ in Eq.~(\ref{Sgen2}). These
superoperators are completely determined by the four Kraus operators 
$\mathsf{K}_j=\Lambda_j \mathsf{U}_\xi$.
In particular, the superoperators $\mathcal{S}_{00}(\rho)$ and $\mathcal{S}_{11}(\rho)$ 
describe the (identical) action on the probe qubit $\rho$ of the standard cascades (1)--(2) and
(2)--(1), expressible as $\mathcal{S}_{00}(\rho)=\mathcal{S}_{11}(\rho) =
\sum_{j=1}^4 \sum_{k=1}^4 \Lambda_j \mathsf{U}_\xi \Lambda_k \mathsf{U}_\xi \rho 
\mathsf{U}_\xi^\dagger \Lambda_k^\dagger \mathsf{U}_\xi^\dagger \Lambda_j^\dagger$,
equivalently performing the Bloch vector transformation
$\vec{r} \mapsto AU_\xi \bigl( AU_\xi \vec{r}+\vec{c} \,\bigr) +\vec{c} $.
Meanwhile, the superoperator $\mathcal{S}_{01}(\rho)$ following from Eq.~(\ref{S01}) with
$\mathsf{K}_j^{(1)}=\mathsf{K}_j^{(2)} =\Lambda_j \mathsf{U}_\xi$, is responsible for the 
distinctive properties stemming from the indefinite superposition of causal orders, as we are going 
to see in more detail.

\section{Measurement for estimation} \label{meas_sec}

We want to exploit the switched channel to perform parameter estimation, chiefly of the phase 
$\xi$ of the unitary $\mathsf{U}_\xi$, but possibly also of the parameters $p$ or $\gamma$ of the 
qubit thermal noise. For this objective, it is possible to measure the two qubits -- probe and 
control -- delivered at the output of the switched channel in the joint state 
$\mathcal{S}(\rho \otimes \rho_c)$ from Eq.~(\ref{Sgen2}).
A simpler approach would be to measure only the probe qubit, since it is the qubit that interacts
with the process $\mathsf{U}_\xi$ and noise $\mathcal{N}(\cdot)$.
If at the same time the control qubit is discarded (unobserved), the measurement on the probe
qubit is ruled by the reduced density operator obtained by partially tracing the joint state 
$\mathcal{S}(\rho \otimes \rho_c)$ of Eq.~(\ref{Sgen2}) over the control qubit. It results as
$\rho^{\rm prob}=\tr_{\rm control}\bigl[ \mathcal{S}(\rho \otimes \rho_c) \bigr] = 
\mathcal{S}_{00}(\rho)$, which is nothing else than the density operator of a single probe qubit
that would have traversed the standard cascade (1)--(2) or (2)--(1), with no effect from the
superposition of causal orders.
An interesting alternative, manifesting the specific and useful properties stemming from the
superposition of causal orders, is to choose to measure the control qubit alone while
discarding the probe qubit. In this circumstance, the measurement on the control
qubit is ruled by the reduced density operator obtained by partially tracing the joint state 
$\mathcal{S}(\rho \otimes \rho_c)$ of Eq.~(\ref{Sgen2}) over the probe qubit. It results as
$\rho^{\rm con}=\tr_{\rm probe}\bigl[ \mathcal{S}(\rho \otimes \rho_c) \bigr]$, yielding when the 
control qubit is initialized in the pure state
$\ket{\psi_c}=\sqrt{p_c}\ket{0_c}+\sqrt{1-p_c}\ket{1_c}$,
\begin{equation}
\rho^{\rm con}=p_c \ket{0_c}\bra{0_c} + (1-p_c) \ket{1_c}\bra{1_c}  + 
Q_c \sqrt{(1-p_c)p_c}\, \Bigl( \ket{1_c}\bra{0_c} + \ket{0_c}\bra{1_c} \Bigr) \;,
\label{Sgenqb_tp2}
\end{equation}
since $\mathcal{S}_{00}(\rho)$ is a qubit density operator giving 
$\tr[ \mathcal{S}_{00}(\rho)] =1$, and where we have defined the factor
\begin{equation}
Q_c =  \tr\bigl[ \mathcal{S}_{01}(\rho) \bigr] \;.
\label{Qfactor1}
\end{equation}
A very interesting feature is that this factor $Q_c$ in Eq.~(\ref{Qfactor1}) is in general
dependent on the properties of both the unitary $\mathsf{U}_\xi$ and the thermal noise 
$\mathcal{N}(\cdot)$, as we are going see in more detail below. This means that the control
qubit alone can be measured at the output of the switched channel, while discarding the probe
qubit, and in this way extract information about the probed processes $\mathsf{U}_\xi$ and 
$\mathcal{N}(\cdot)$. This is remarkable since the control qubit does not interact with the
probed processes $\mathsf{U}_\xi$ and $\mathcal{N}(\cdot)$, only the probe qubit does so.
Nevertheless, the type of coupling induced by the switched channel with indefinite causal order,
as conveyed by Eq.~(\ref{Sgen2}), transfers information from the probed processes 
$\mathsf{U}_\xi$ and $\mathcal{N}(\cdot)$ to the control qubit.

To quantify the performance of the control qubit, for a task of parameter estimation, a generally 
meaningful criterion is the quantum Fisher information, contained in the state $\rho^{\rm con}$, 
about some parameter of interest probed in the switched channel 
\cite{Barndorff00,Paris09,Facchi10,Chapeau17}. The quantum Fisher information stands as an upper 
bound to the classical Fisher information, which in turn determines the smallest mean-squared 
error that can be achieved in the estimation. In this way, the quantum Fisher information is a 
fundamental criterion characterizing the best performance that can be envisaged, applying equally 
to any estimation strategies, and in this respect dispensing to refer to an explicit quantum 
measurement and an explicit estimator. We will concentrate here on estimating the phase $\xi$ of 
the noisy unitary process $\mathsf{U}_\xi$, which is often a parameter of prime interest in quantum 
metrology. Nevertheless, a comparable analysis would hold equally for estimating other parameters, 
such as the probability $p$ or the damping factor $\gamma$ of the thermal noise. Based on 
\cite{Chapeau21}, when measuring the control qubit of the switched channel in the state
$\rho^{\rm con}$ of Eq.~(\ref{Sgenqb_tp2}), the quantum Fisher information for estimating the phase 
$\xi$ follows as
\begin{equation}
F_q^{\rm con}(\xi) = 4(1-p_c)p_c\dfrac{\bigl[\partial_\xi Q_c(\xi) \bigr]^2}{1-Q_c^2(\xi)}  \;,
\label{Fq_con1_g}
\end{equation}
involving the derivative $\partial_\xi Q_c(\xi) \equiv \partial Q_c(\xi)/ \partial\xi$ of the 
factor $Q_c \equiv Q_c(\xi)$ from Eq.~(\ref{Qfactor1}) seen as a function of $\xi$ and 
characterizing the measured state $\rho^{\rm con}$ of Eq.~(\ref{Sgenqb_tp2}). This quantum Fisher 
information in Eq.~(\ref{Fq_con1_g}) is maximized for a control qubit prepared with the probability 
$p_c =1/2$, i.e.\ with a maximally indefinite order via an even superposition of the two orders 
(1)--(2) and (2)--(1) in Fig.~\ref{figSwi1}, so as to give at $p_c =1/2$, 
\begin{equation}
F_q^{\rm con}(\xi) = \dfrac{\bigl[\partial_\xi Q_c(\xi) \bigr]^2}{1-Q_c^2(\xi)}  \;.
\label{Fq_con1}
\end{equation}
We essentially focus the analysis on this optimal configuration at $p_c=1/2$ in the sequel.

\bigbreak
For a meaningful example that illustrates essential distinctive properties accessible with the 
control qubit of the switched channel, while remaining analytically tractable with closed-form 
expressions of moderate size to handle, we consider the situation of a unitary $\mathsf{U}_\xi$
in Eq.~(\ref{Uxi}) with axis $\vec{n}=[0, 0, 1]^\top =\vec{e}_z$. The four Kraus operators 
$\mathsf{K}_j=\Lambda_j \mathsf{U}_\xi$ then follow to determine the two-qubit joint state 
$\mathcal{S}(\rho \otimes \rho_c)$ of Eq.~(\ref{Sgen2}), which is explicitly worked out in the 
Appendix. We also obtain in the Appendix the characterization of the operator 
$\mathcal{S}_{01}(\rho)$ whose trace is computed according to Eq.~(\ref{Qfactor1}) to yield
\begin{equation}
Q_c(\xi) = 2\gamma \sqrt{1-\gamma}\,\Bigl[(1-2p)\rho_{00}+p \Bigr] \cos(\xi)
+ (2-\gamma)\gamma (2p-1) \rho_{00} + (1-\gamma p)^2 \;.
\label{Qxi1}
\end{equation}
The term $\rho_{00} \in [0, 1]$ in Eq.~(\ref{Qxi1}) conveys the influence from the initial 
preparation in Eq.~(\ref{ro_in1}) of the input probe qubit. Follows the derivative
\begin{equation}
\partial_\xi Q_c(\xi) = -2\gamma \sqrt{1-\gamma}\,\Bigl[(1-2p)\rho_{00}+p \Bigr] \sin(\xi) \;,
\label{dQxi1}
\end{equation}
providing a complete determination of the quantum Fisher information $F_q^\text{con}(\xi)$ of
Eq.~(\ref{Fq_con1}).

For comparison, a meaningful reference is the quantum Fisher information $F_q(\xi)$ accessible for 
estimating the phase $\xi$ when directly measuring the output qubit of a single standard channel 
as in Fig.~\ref{fig_blocUN}. For an input probe qubit with Bloch vector $\vec{r}$ as in 
Eq.~(\ref{ro_in1}), we have for instance from \cite{Chapeau16} the expression
\begin{equation}
F_q(\xi) = \dfrac{\bigl[(AU_\xi\vec{r}+\vec{c}\,) A(\vec{n}\times U_\xi\vec{r}\,)\bigr]^2}
{1-(AU_\xi\vec{r}+\vec{c}\,)^2} +
\bigl[ A(\vec{n}\times U_\xi\vec{r}\,) \bigr]^2 \;.
\label{Fq_ref}
\end{equation}
The conditions for maximizing the quantum Fisher information $F_q(\xi)$ of Eq.~(\ref{Fq_ref}) are
analyzed for instance in \cite{Chapeau15,Chapeau16}.
With the thermal noise model of Eqs.~(\ref{4.tgad1})--(\ref{4.A_GAD}), the quantum Fisher 
information $F_q(\xi)$ of Eq.~(\ref{Fq_ref}) for the standard channel of Fig.~\ref{fig_blocUN}, can 
reach the overall maximum value $F_q^\text{max}(\xi)=1-\gamma$ when three conditions are 
satisfied \cite{Chapeau15,Chapeau16}: 
(i) the input probe must be pure with $\| \vec{r}\, \|=1$; 
(ii) the Bloch vector $\vec{r}$ defining the input probe must be orthogonal 
to the rotation axis $\vec{n}$ of the unitary $\mathsf{U}_\xi$ under estimation;
(iii) the vector $\vec{n}\times U_\xi\vec{r}$ in Eq.~(\ref{Fq_ref}) must be orthogonal to the
$Oz$ axis of $\mathbbm{R}^3$ set by the thermal noise of Eq.~(\ref{4.A_GAD}).
This usually implies a $\xi$-dependent condition on the rotated probe $U_\xi\vec{r}$,
that usually cannot be met by a fixed input probe $\vec{r}$, but can be circumvented by adaptively 
adjusting the input probe in an iterative estimation protocol \cite{Brivio10,Okamoto12}.
When gradually departing from these conditions, the Fisher information $F_q(\xi)$ gradually decays 
below the maximum $F_q^\text{max}(\xi)=1-\gamma$. Especially, when the input probe $\vec{r}$ tends to 
align with the axis $\vec{n}$, or when it depolarizes as $\| \vec{r}\, \| \rightarrow 0$, then 
$F_q(\xi)$ goes to zero and phase estimation from the standard channel of Fig.~\ref{fig_blocUN} 
becomes completely inoperative. Standard phase estimation resting on the same elementary 
probe--unitary interaction $\rho \mapsto \mathsf{U}_\xi \rho \mathsf{U}_\xi^\dagger$ will remain 
inoperative in the same conditions, even with more elaborate scenarios as with several passes through 
$\mathsf{U}_\xi$ of the probe or several probing qubits possibly entangled 
\cite{Demkowicz14,Chapeau21}.

By contrast, a distinctive feature is that the control qubit of the switched channel exhibits a
Fisher information $F_q^{\rm con}(\xi)$ in Eq.~(\ref{Fq_con1}) not limited in the same way by the 
conditions (i)--(iii) above. In this respect, the control qubit gives access to novel 
capabilities relevant to phase estimation and complementary to those offered by the standard 
approach of Eq.~(\ref{Fq_ref}). Especially, the performance of the control qubit for phase 
estimation, as assessed by the Fisher information $F_q^{\rm con}(\xi)$ of Eq.~(\ref{Fq_con1}), is 
not adversely impacted by ill configurations of the input probe $\vec{r}$ of Eq.~(\ref{ro_in1}) in 
relation to the rotation axis $\vec{n}$. Even when $\vec{r} \varparallel \vec{n}$ with an input 
probe $\vec{r}$ parallel to the rotation axis $\vec{n}$, or when $\| \vec{r}\, \| =\vec{0}$ with a 
fully depolarized input probe, the control qubit of the switched channel remains operative for phase 
estimation. We specially focus in the sequel on analyzing the performance of the control qubit for 
phase estimation in these conditions where standard phase estimation becomes completely inoperative, 
with $F_q(\xi) \equiv 0$ in Eq.~(\ref{Fq_ref}).

\begin{figure}[htb]
\centerline{\includegraphics[width=94mm]{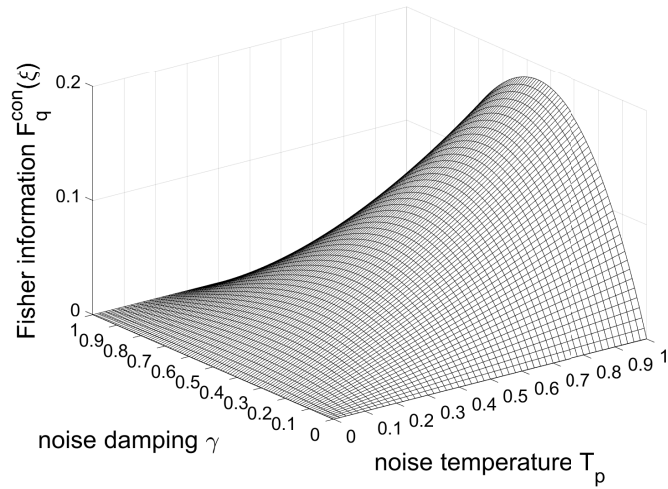}}
\caption[what appears in lof LL p177]
{For the control qubit of the switched channel, quantum Fisher information $F_q^{\rm con}(\xi)$ 
of Eq.~(\ref{Fq_con1}), for a phase $\xi =\pi /4$, as a function of the effective temperature 
$T_p=2(1-p)$ and damping factor $\gamma$ of the quantum thermal noise of
Eqs.~(\ref{4.tgad1})--(\ref{4.A_GAD}).
The input probe is prepared in the pure state $\rho =\ket{0}\bra{0}$ with Bloch vector 
$\vec{r}=\vec{e}_z$ parallel to the rotation axis $\vec{n}=\vec{e}_z$ of the unitary 
$\mathsf{U}_\xi$, making standard phase estimation completely inoperative.
}
\label{figFqT1}
\end{figure}

\begin{figure}[htb]
\centerline{\includegraphics[width=94mm]{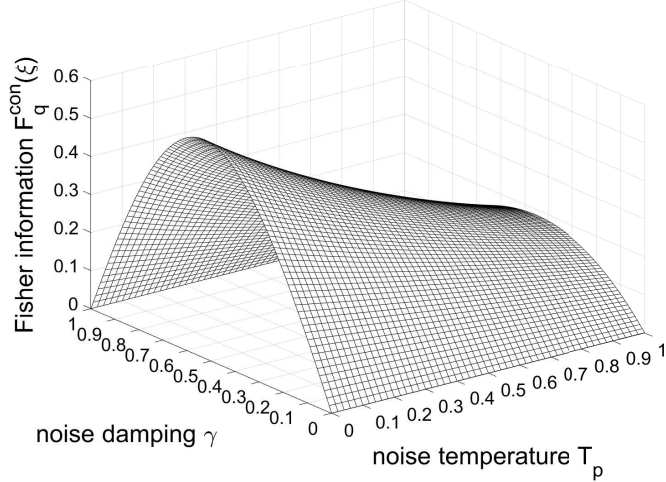}}
\caption[what appears in lof LL p177]
{Same as in Fig.~\ref{figFqT1}, except that the input probe is prepared 
in the pure state $\rho =\ket{1}\bra{1}$ with Bloch vector 
$\vec{r}=-\vec{e}_z$ parallel to the rotation axis $\vec{n}=\vec{e}_z$ of the unitary 
$\mathsf{U}_\xi$, also making standard phase estimation completely inoperative.
}
\label{figFqT2}
\end{figure}

\begin{figure}[htb]
\centerline{\includegraphics[width=94mm]{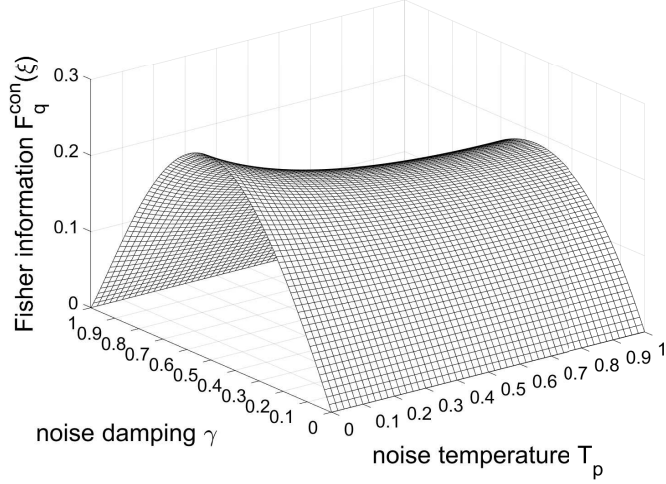}}
\caption[what appears in lof LL p177]
{Same as in Fig.~\ref{figFqT1}, except that the input probe is prepared in the fully depolarized
mixed state $\rho =\mathrm{I}_2 /2$ with Bloch vector $\vec{r}=\vec{0}$, where standard phase 
estimation is also completely inoperative.
}
\label{figFqT3}
\end{figure}

\begin{figure}[htb]
\centerline{\includegraphics[width=84mm]{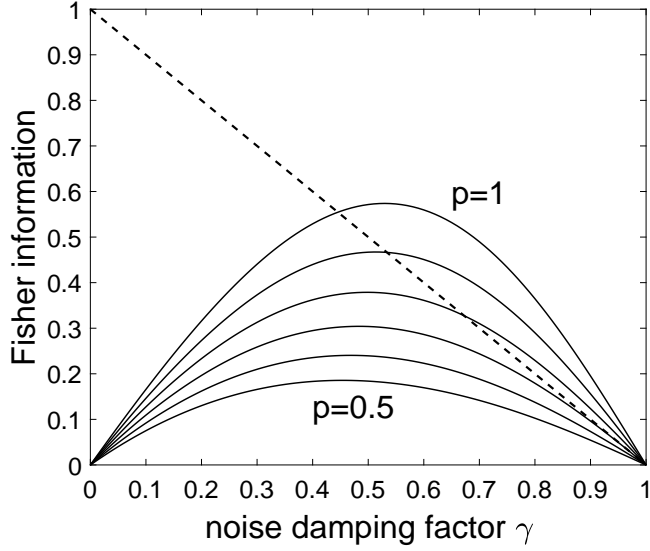}}
\caption[what appears in lof LL p177]
{Solid lines: in the conditions of Fig.~\ref{figFqT2}, the quantum Fisher information 
$F_q^{\rm con}(\xi)$ of Eq.~(\ref{Fq_con1}) as a function of the noise damping factor $\gamma$,
for 6 values of the probability $p$ fixed by the noise temperature: 
the uppermost curve is for $p=1$ at zero temperature, then in descending order of the curves are
$p=0.9$, $0.8$, $0.7$ and $0.6$, finally the lowermost curve is for $p=0.5$ at infinite temperature.
Dashed curve: maximal quantum Fisher information $F_q^\text{max}(\xi)=1-\gamma$ accessible in 
Eq.~(\ref{Fq_ref}) with the standard channel of Fig.~\ref{fig_blocUN} operated in its optimal
conditions with an input probe prepared in the pure state $\ket{+}$ with Bloch vector 
$\vec{r}=\vec{e}_x$.
}
\label{figFqGa1}
\end{figure}

Figures~\ref{figFqT1} to \ref{figFqGa1} illustrate the impact of the thermal noise parameters on 
the performance $F_q^{\rm con}(\xi)$ of the control qubit of the switched channel for phase 
estimation, in conditions where standard phase estimation is completely inoperative.

Figures~\ref{figFqT1} and \ref{figFqT2} deal with two cases of a pure input probe,
$\rho =\ket{0}\bra{0}$ associated with $\rho_{00} =1$ in Fig.~\ref{figFqT1}, and
$\rho =\ket{1}\bra{1}$ associated with $\rho_{00} =0$ in Fig.~\ref{figFqT2}. In both cases
the input Bloch vector $\vec{r}$ is parallel to the rotation axis $\vec{n}=\vec{e}_z$.
In this circumstance, for the standard estimation the Fisher information $F_q(\xi)$ of 
Eq.~(\ref{Fq_ref}) vanishes since $U_\xi \vec{r} \varparallel \vec{n}$, manifesting the
impossibility of estimating the phase by measuring the probe qubit. By contrast, the control 
qubit of the switched channel remains efficient for estimation, as manifested by a nonvanishing 
Fisher information $F_q^{\rm con}(\xi)$ in Figs.~\ref{figFqT1}--\ref{figFqT2}. 

In addition, Figs.~\ref{figFqT1}--\ref{figFqT2} illustrate two typical behaviors accessible to the 
Fisher information $F_q^{\rm con}(\xi)$ of Eq.~(\ref{Fq_con1}), and therefrom to the estimation 
efficiency, upon increasing the temperature of the thermal bath, at any fixed value of the damping 
factor $\gamma$. Depending on the input probe $\rho$, Fig.~\ref{figFqT1} shows the possibility of a
Fisher information $F_q^{\rm con}(\xi)$ that increases as the noise temperature $T_p$ increases,
while Fig.~\ref{figFqT2} shows a decreasing $F_q^{\rm con}(\xi)$. 
This manifests that the (thermal) noise is not univocally detrimental to the estimation efficiency 
from the switched channel, as also observed in \cite{Chapeau21} with depolarizing noise. 
Increasing the level of noise, via increasing the noise temperature in Fig.~\ref{figFqT1}, may
enhance the Fisher information $F_q^{\rm con}(\xi)$ and therefore the efficiency of the control
qubit for the phase estimation.
Such a versatile role of noise in the switched channel can also be substantiated in the following way.
In the elementary noisy unitary channel of 
Fig.~\ref{fig_blocUN}, the noise $\mathcal{N}(\cdot)$ has the natural effect of degrading the 
probe qubit, and hence the estimation efficiency. Yet, when two such channels are superposed as in 
Fig.~\ref{figSwi1}, the noise is necessary to make the two superposed channels (1) and (2) 
distinguishable, and to couple the control qubit to the unitary $\mathsf{U}_\xi$. At vanishing noise, 
the two channels (1) and (2) are two identical unitary channels with one single Kraus operator 
$\mathsf{K}_1=\mathsf{U}_\xi$, so that the joint probe-control output state of Eq.~(\ref{Sgen2}) 
reduces to the separable state
$\mathcal{S}(\rho \otimes \rho_c)=\mathcal{S}_{00}(\rho) \otimes \ket{\psi_c}\bra{\psi_c}$,
and the control qubit does not couple to the unitary $\mathsf{U}_\xi$. This in particular 
entails, at vanishing noise, when the damping $\gamma \rightarrow 0$ in Eq.~(\ref{4.A_GAD}), 
a coupling factor $Q_c(\xi)$ in Eq.~(\ref{Qxi1}) which becomes independent of the phase $\xi$, so 
that the control qubit cannot serve to estimate the phase $\xi$. A nonvanishing amount of noise is 
required to couple the control qubit to the phase $\xi$ and maintain a $\xi$-dependent $Q_c(\xi)$ 
in Eq.~(\ref{Qxi1}). Increasing the amount of noise, in this way, may be beneficial to the efficiency
of the control qubit for estimation, depending on the conditions, as determined by the expression
of Eq.~(\ref{Fq_con1}) for the quantum Fisher information $F_q^{\rm con}(\xi)$.

Figure~\ref{figFqT3} shows the situation of a fully depolarized input probe $\rho =\mathrm{I}_2 /2$ 
with $\rho_{00}=1/2$ and Bloch vector $\vec{r}=\vec{0}$, which maintains a nonvanishing Fisher 
information $F_q^{\rm con}(\xi)$ in Eq.~(\ref{Fq_con1}) and therefore the capability of the control 
qubit of the switched channel for estimating the phase $\xi$. On the contrary, with the standard 
channel of Fig.~\ref{fig_blocUN}, a fully depolarized input probe $\rho =\mathrm{I}_2 /2$ on the 
unitary, undergoes the transformation 
$\rho =\mathrm{I}_2 /2 \mapsto \mathsf{U}_\xi \rho \mathsf{U}_\xi^\dagger =\mathrm{I}_2 /2$,
and extracts no information about $\mathsf{U}_\xi$, so that it is completely inoperative for
estimating its phase $\xi$.

Figures~\ref{figFqT1}--\ref{figFqT3} also clearly show a non-monotonic action of the damping 
factor $\gamma =1-e^{-t/\tau_1}$ on the quantum Fisher information $F_q^{\rm con}(\xi)$, at any 
fixed temperature of the thermal noise.
A damping factor $\gamma =1-e^{-t/\tau_1}$ tending to zero at an extremely brief exposition time 
$t \ll \tau_1$ to the noise, is the situation of a vanishing thermal noise. 
As explained above, at vanishing noise the control qubit is no longer coupled to the unitary 
$\mathsf{U}_\xi$ in the switched channel, whence the vanishing Fisher information 
$F_q^{\rm con}(\xi)$ at $\gamma =0$ in Figs.~\ref{figFqT1}--\ref{figFqT3}. 
Also, when $\gamma =1-e^{-t/\tau_1}\rightarrow 1$, with an increasing exposition time 
$t \gg \tau_1$, the Fisher information $F_q^{\rm con}(\xi)$ also tends to vanish, as 
visible in Figs.~\ref{figFqT1}--\ref{figFqT3}.
This points to an intermediate damping $\gamma$, i.e.\ an intermediate exposition time $t$ to the 
noise, to maximize the estimation efficiency $F_q^{\rm con}(\xi)$, with $t$ not too short so that 
the control qubit sufficiently couples to $\mathsf{U}_\xi$, and $t$ not too long to avoid 
thermalization where the process terminates in an equilibrium state independent of 
$\mathsf{U}_\xi$. This is reflected in a Fisher information $F_q^{\rm con}(\xi)$ of the control 
qubit, culminating at a maximum for an intermediate damping $\gamma$, whose precise value is 
slightly dependent on the temperature of the thermal bath, as observed in the finer view
provided by Fig.~\ref{figFqGa1} in the illustrative conditions of Fig.~\ref{figFqT2}.

\medbreak 

Such a constructive role of the thermal noise here, which may benefit to the efficiency of the
control qubit for estimation, was similarly observed with the depolarizing noise in the coherently 
superposed channels of \cite{Chapeau21,Chapeau21b}.
Such a constructive role of noise can be related to a phenomenon of stochastic resonance,
a general effect taking place in diverse information processing operations,
classical \cite{Gammaitoni98,Chapeau99c,McDonnell08,Duan14} or quantum 
\cite{Ting99,Bowen06,Chapeau15c,Gillard17,Gillard19}, and where maximum efficacy is observed 
at a non-vanishing optimal amount of noise. Here it confirms with a novel scenario that quantum 
noise or decoherence can sometimes turn beneficial to quantum information processing.

Figure~\ref{figFqGa1} also compares the quantum Fisher information of the 
control qubit of the switched channel and that of the standard channel of Fig.~\ref{fig_blocUN} 
when both channels are operated at their best.
When the standard channel of Fig.~\ref{fig_blocUN} can be operated in its optimal conditions,
matching the three conditions (i)--(iii) mentioned above after Eq.~(\ref{Fq_ref}), the
maximal quantum Fisher information $F_q^\text{max}(\xi)=1-\gamma$ it obtains is usually
superior to $F_q^{\rm con}(\xi)$ from the switched channel, over a significant range of the
noise parameters $(p, \gamma)$. Nevertheless, as visible in Fig.~\ref{figFqGa1}, there exists a
range, at high $p$ and $\gamma$, where $F_q^{\rm con}(\xi)$ from the switched channel becomes
superior to $F_q^\text{max}(\xi)=1-\gamma$ from the standard channel.
In the conditions of Fig.~\ref{figFqGa1}, as $p$ and $\gamma$ go to $1$,
one has for the control qubit the Fisher information $F_q^{\rm con}(\xi) \rightarrow 2(1-\gamma)$.
The same quantum Fisher information $2(1-\gamma)=2F_q^\text{max}(\xi)$ could be reached by two
independent qubits traversing the standard channel of Fig.~\ref{fig_blocUN} and then being
measured. A single qubit traversing twice, in two passes, the standard channel of 
Fig.~\ref{fig_blocUN}, would achieve a quantum Fisher information comparable to Eq.~(\ref{Fq_ref})
but acting on the Bloch vector $AU_\xi\bigl(AU_\xi \vec{r}+\vec{c}\bigr)+\vec{c}$ as transformed
by the two passes. Yet this would lead to a poorer performance compared with the two independent 
qubits, because when starting the second pass the (noisy) qubit would no longer be in the optimal 
input state. Two entangled probe qubits, although more complicated to handle, could even be 
envisaged to probe the standard channel of Fig.~\ref{fig_blocUN}, to benefit from the so-called 
Heisenberg enhanced performance, but the optimal configurations in the presence of thermal noise are 
not fully characterized, and it is known that the Heisenberg enhanced performance is very fragile to 
noise \cite{Giovannetti11,Demkowicz12,Demkowicz14,Chapeau18}. In the high range of the noise 
parameters $(p, \gamma)$ in Fig.~\ref{figFqGa1}, the control qubit of the switched channel performs 
as well as two independent qubits across the standard channel, but it can be measured alone. The two 
channels compared in Fig.~\ref{figFqGa1} are rather distinct in their constitution and mode of 
operation, means and resources they imply. The standard channel which is more directly intended for 
estimation, however, does not always achieve the best performance of the two channels.
The two channels are rather complementary, especially as the switched channel offers estimation
capabilities inaccessible to the standard channel and forming the main focus of this study.

\begin{figure}[htb]
\centerline{\includegraphics[width=84mm]{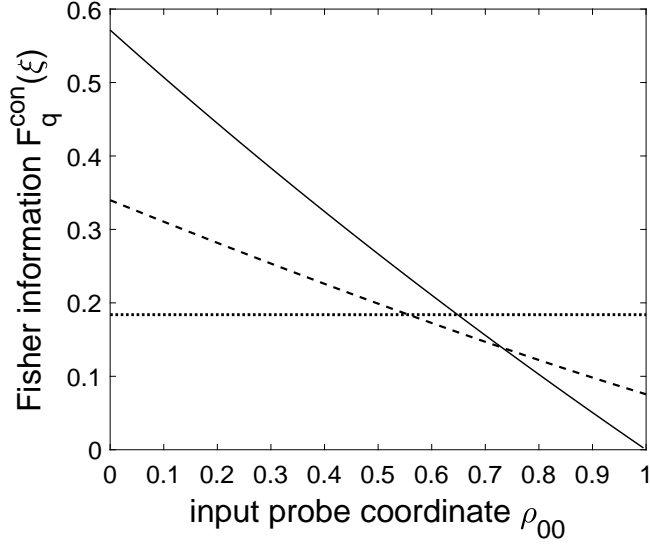}}
\caption[what appears in lof LL p177]
{Quantum Fisher information $F_q^{\rm con}(\xi)$ of Eq.~(\ref{Fq_con1}), for a phase $\xi =\pi /4$,
as a function of the input probe coordinate $\rho_{00}$, with a noise damping factor 
$\gamma =0.5$ and a noise probability $p=1$ corresponding to the zero temperature $T=0$ 
(solid line), $p=0.75$ at intermediate temperature $T$ (dashed line), $p=0.5$ at infinite 
temperature $T=\infty$ (dotted line). The pure input probe $\rho=\ket{1}\bra{1}$ is at
$\rho_{00} =0$, while $\rho=\ket{0}\bra{0}$ is at $\rho_{00} =1$, the fully depolarized input 
probe $\rho =\mathrm{I}_2 /2$ is at $\rho_{00} =1/2$, all three configurations where standard 
phase estimation is completely inoperative.
}
\label{figFqR0}
\end{figure}

Figure~\ref{figFqR0} shows typical evolutions of the quantum Fisher information 
$F_q^{\rm con}(\xi)$ resulting from Eq.~(\ref{Fq_con1}) and Eqs.~(\ref{Qxi1})--(\ref{dQxi1}) for 
the control qubit of the switched channel, as a function of the coordinate $\rho_{00}$ of the 
input probe $\rho$. 

\smallbreak

The configurations in Fig.~\ref{figFqR0} encompass the pure input probes $\rho=\ket{1}\bra{1}$ at
$\rho_{00} =0$ and $\rho=\ket{0}\bra{0}$ at $\rho_{00} =1$, the fully depolarized input 
probe $\rho =\mathrm{I}_2 /2$ at $\rho_{00} =1/2$, all three configurations where standard 
phase estimation is completely inoperative. Meanwhile, in these configurations, and over the whole
range of $\rho_{00} \in [0, 1]$, the control qubit of the switched channel generally remains 
operative for estimating the phase $\xi$, as established by the nonvanishing Fisher information
$F_q^{\rm con}(\xi)$ in Fig.~\ref{figFqR0}. When seen as a function of $\rho_{00}$, the Fisher 
information $F_q^{\rm con}(\xi)$ resulting from Eqs.~(\ref{Fq_con1})--(\ref{dQxi1}), is a 
decreasing function of $\rho_{00} \in [0, 1]$, so that $F_q^{\rm con}(\xi)$ is always maximized in 
$\rho_{00} =0$, that is, by the pure input probe $\rho=\ket{1}\bra{1}$, as in 
Figs.~\ref{figFqT2} and \ref{figFqGa1}.

\bigbreak
Finally, it is possible to refer to an explicit measurement of the control qubit in the state
$\rho^{\rm con}$ of Eq.~(\ref{Sgenqb_tp2}), by means of a von Neumann measurement in the Hadamard 
basis $\bigl\{\ket{+}, \ket{-} \bigr\}$. The two measurement outcomes of projecting the control 
qubit on $\ket{+}$ or $\ket{-}$ occur with the probabilities 
$\Pr\bigl\{\ket{\pm}\bigr\}=P^{\rm con}_\pm$ which are
\begin{equation}
P^{\rm con}_\pm =\braket{\pm | \rho^{\rm con} | \pm}=
\frac{1}{2} \pm \sqrt{(1-p_c)p_c} Q_c \;.
\label{Pc+a}
\end{equation}
This binary measurement result contains, about the unknown phase $\xi$, the classical Fisher
information \cite{Chapeau16} 
\begin{eqnarray}
\label{Fc2}
F_c^{\rm con}(\xi) &=& \dfrac{(\partial_\xi P^{\rm con}_+)^2}{(1-P^{\rm con}_+)P^{\rm con}_+} \\
\label{Fc2_b}
&=& \dfrac{4(1-p_c)p_c\bigl[\partial_\xi Q_c(\xi) \bigr]^2}{1-4(1-p_c)p_c Q_c^2(\xi)} \;.
\end{eqnarray}
At the optimal preparation $p_c=1/2$ of the control qubit, $F_c^{\rm con}(\xi)$ 
gets maximized at the level $F_q^{\rm con}(\xi)$ of
Eq.~(\ref{Fq_con1}). This establishes the measurement of the control qubit in the Hadamard basis 
$\bigl\{\ket{+}, \ket{-} \bigr\}$ as an optimal measurement, reaching the highest possible
classical Fisher information $F_c^{\rm con}(\xi)=F_q^{\rm con}(\xi)$, and by means of the maximum 
likelihood estimator reaching the smallest mean-squared estimation error.
Measurement in the Hadamard basis $\bigl\{\ket{+}, \ket{-}\bigr\}$ is equivalent to measuring the 
spin observable $\vec{\omega}_c \cdot \vec{\sigma}$ characterized by the measurement vector 
$\vec{\omega}_c =[1, 0, 0]^\top =\vec{e}_x$. It is remarkable that such a fixed optimal measurement
reaching $F_c^{\rm con}(\xi)=F_q^{\rm con}(\xi)$
exists for the control qubit, independent of the axis $\vec{n}$ and angle $\xi$ of the unitary 
$\mathsf{U}_\xi$ under estimation. By comparison, standard estimation approaches ruled by 
Eq.~(\ref{Fq_ref}) are in general not granted with such a fixed optimal measurement, and to reach 
the optimum of $F_c(\xi)=F_q(\xi)$ matching the classical and quantum Fisher informations, they 
usually need \cite{Chapeau16} to measure a spin observable $\vec{\omega}\cdot \vec{\sigma}$ with a 
measurement vector $\vec{\omega}$ dependent on $\vec{n}$ and $\xi$, which is not generally feasible 
with an unknown phase $\xi$.

In the switched channel, the measurement in the Hadamard basis $\bigl\{\ket{+}, \ket{-}\bigr\}$ of 
the control qubit leaves the probe qubit in the unnormalized conditional state
\begin{equation}
\rho_\pm = {}_c\langle \pm | \mathcal{S}(\rho \otimes \rho_c) | \pm \rangle_c
=\dfrac{1}{2}\mathcal{S}_{00}(\rho) \pm \sqrt{(1-p_c)p_c} \,\mathcal{S}_{01}(\rho) \;.
\label{-S+}
\end{equation}
After proper normalization by the corresponding probabilities of occurrence 
$P^{\rm con}_\pm =\tr(\rho_\pm)$ of Eq.~(\ref{Pc+a}), one obtains the post-measurement state of the 
probe qubit $\rho_\pm^{\rm post}=\rho_\pm /P^{\rm con}_\pm$ conditioned on the measurement outcome 
obtained on the control qubit. It is observed that in general this state $\rho_\pm^{\rm post}$ of 
the probe is dependent on the phase $\xi$ under estimation, so that the probe qubit can be 
subsequently measured, after the control qubit, so as to extract additional information about the 
phase $\xi$. Proceeding in this way is a sequential strategy, implementing two separable 
measurements, successively on the control and probe qubits. A joint entangled measurement of the 
qubit pair could also be envisaged, based on the explicit characterization of the joint state 
$\mathcal{S}(\rho \otimes \rho_c)$ developed in the Appendix. 
Such two-qubit strategies are a priori more complicated to analyze, to optimize and to implement. 
Interesting properties could nevertheless result, based on the non-standard capabilities contributed 
by the control qubit we reported here. 
This direction is open for further investigation to complement the present demonstration of 
non-standard capabilities offered by the control qubit of the switched channel for quantum phase 
estimation in presence of thermal noise.

\section{Discussion and conclusion}

We have considered the switched quantum channel with indefinite causal order of Fig.~\ref{figSwi1} 
when the elementary channels (1) and (2) are two copies of the noisy unitary qubit channel of 
Fig.~\ref{fig_blocUN}. From the characterization of the two-qubit probe-control state 
$\mathcal{S}(\rho \otimes \rho_c)$ from Eq.~(\ref{Sgen2}) delivered by the switched channel, we have 
specifically investigated the properties accessible for the fundamental metrological task of phase 
estimation on the unitary process $\mathsf{U}_\xi$ in the presence of a quantum thermal noise 
$\mathcal{N}(\cdot)$. This study complements the report of \cite{Chapeau21} that concentrated on 
the qubit depolarizing noise, and it brings further results useful to a broader appreciation of 
the capabilities of switched channels with indefinite causal order for quantum metrology.
We have observed here, as in \cite{Chapeau21}, that the noise is an important ingredient to make 
the elementary channels (1) and (2) distinguishable, so that the superposition of 
Fig.~\ref{figSwi1} does not reduce to a standard cascade of two identical unitaries 
$\mathsf{U}_\xi$ and so that it can exhibit novel properties inaccessible to standard cascades 
with definite causal order.

Another important property is that in the presence of noise, the control qubit of the switched
channel gets coupled to the unitary $\mathsf{U}_\xi$, in such a way that the control qubit alone, 
although it does not actively interact with the unitary $\mathsf{U}_\xi$, can be measured to 
estimate its phase $\xi$. In standard estimation approaches, such noninteracting qubits, to be of 
some use, need to be jointly measured with the active probing qubits \cite{Demkowicz14,Chapeau17}.
This points here to non-standard quantum correlations induced in the switched channel with 
indefinite causal order, offering non-standard properties useful to quantum estimation, as also
observed in \cite{Chapeau21}. We then have concentrated the analysis on such non-standard properties 
specific to the switched channel. Especially, we have analyzed phase estimation from the control 
qubit alone in configurations of the input probe where standard estimation approaches become 
completely inoperative, even when repeated with multiple probing qubits or several passes across 
the unitary $\mathsf{U}_\xi$ under estimation. In particular, we have shown that the control qubit 
remains efficient for phase estimation even with an input probe $\vec{r}$ aligned with the rotation 
axis $\vec{n}$ of the unitary $\mathsf{U}_\xi$, or with a fully depolarized input probe.

In the present study, the probe and control qubits have different status and position, with a probe 
qubit that directly interacts with the noisy unitary $\mathsf{U}_\xi$ and a control qubit that does 
not. This conforms with the common reference framework for studying switched non-causal order, of 
two noisy channels whose causal order is driven by a noise-free control signal, as for instance in 
\cite{Ebler18,Procopio19,Procopio20,Loizeau20,Chiribella12,Koudia19,Frey19,Mukhopadhyay18,Chapeau21}.
However, when the control qubit is also affected by some quantum noise or alteration, the present 
analysis can be used to predict how the essential properties stemming from non-causal order, 
as conveyed by the quantum Fisher information $F_q^{\rm con}(\xi)$, are preserved or gradually 
diminished. When the input state $\rho_c$ of the control qubit is a pure state that gradually 
departs from the optimum pure state at $p_c=1/2$, the quantum Fisher information 
$F_q^{\rm con}(\xi)$ in Eq.~(\ref{Fq_con1_g}) of the control qubit gradually decays, but does not 
vanish until the superposition of orders completely disappears at $p_c=0$ or $p_c=1$. When the 
control state $\rho_c$ becomes a mixed state, Eq.~(\ref{Sgen2}) shows that its action is still 
operative and conveyed by its matrix elements in the basis $\bigl\{\ket{0_c}, \ket{1_c} \bigr\}$.
When tracing Eq.~(\ref{Sgen2}) over the probe, the matrix elements $\braket{0_c|\rho_c|1_c}$ and 
$\braket{1_c|\rho_c|0_c}=\braket{0_c|\rho_c|1_c}^*$ are transported as multiplicative factors on 
$Q_c$ in Eq.~(\ref{Sgenqb_tp2}), in place of $\sqrt{(1-p_c)p_c}$ of the pure control. The factor 
$Q_c \equiv Q_c(\xi)=\tr\bigl[ \mathcal{S}_{01}(\rho) \bigr]$ remains the essential element that 
conveys the dependence on the phase $\xi$ of the transformed state $\rho^{\rm con}$ of the control 
qubit, via Eqs.~(\ref{Qfactor1}) or (\ref{Qxi1}) which are 
unchanged as determined by the probe qubit. As long as $\braket{0_c|\rho_c|1_c}$ is not reduced to 
zero, the control state $\rho^{\rm con}$ of Eq.~(\ref{Sgenqb_tp2}) remains dependent, via
$Q_c(\xi)$, on the phase $\xi$ and can serve to its estimation. This is reflected in a quantum 
Fisher information $F_q^{\rm con}(\xi)$ of the control qubit that gradually decays with 
$|\braket{0_c|\rho_c|1_c}|$, to vanish when $|\braket{0_c|\rho_c|1_c}|$ reaches zero, but not 
earlier. As long as $\rho^{\rm con}$ remains dependent on $\xi$ it can be measured to estimate
$\xi$, and an additional noise altering $\rho^{\rm con}$ before it can be measured, would still 
preserve its capability for phase estimation, until the off-diagonal matrix elements of the control 
qubit state that carry $Q_c(\xi)$ are completely canceled by the noise. In this way, the properties 
of the control qubit for phase estimation are robustly preserved.

Other explorations can be envisaged of the properties of the switched channel with indefinite
causal order, based on the general characterization of the two-qubit output state 
$\mathcal{S}(\rho \otimes \rho_c)$ of Eq.~(\ref{Sgen2}). For instance, estimation of the
thermal noise parameters $\gamma$ or $p$ can be envisaged, and this can again be performed
by measuring the control qubit alone, while discarding the probe qubit interacting with the
thermal bath. This is feasible thanks to the coupling factor $Q_c$ in Eq.~(\ref{Sgenqb_tp2}) 
which bears dependence, as indicated by Eq.~(\ref{Qxi1}), also on the noise parameters
$\gamma$ and $p$ and thus extract information about them. This is still true for instance with
a fully depolarized input probe $\rho =\mathrm{I}_2 /2$, having $\rho_{00} =1/2$. 
Our analysis, in the special case where the phase of the unitary $\mathsf{U}_\xi$ is canceled as 
$\xi \equiv 0$, describes a pure thermal noise channel engaged in a superposition of causal orders, 
as studied for quantum thermometry in \cite{Mukhopadhyay18} in the thermalized regime when the 
damping $\gamma \rightarrow 1$.
Another direction of exploration could be the investigation of extensions and generalizations 
proposed of structures with non-causal order \cite{Araujo14b,Procopio19,Kristjansson20} for 
estimation tasks with noise as we considered here in the quantum switch superposing two causal 
orders.

In this way, the present study brings additional elements and results to better appreciate the 
capabilities of switched quantum channels with indefinite causal order, especially with novel and 
specific properties, to contribute to quantum estimation and quantum metrology and more broadly to 
quantum signal and information processing.

\section*{Appendix A}

\renewcommand{\theequation}{A-\arabic{equation}} 
\setcounter{equation}{0}  

In this Appendix, we explicitly work out the joint state $\mathcal{S}(\rho \otimes \rho_c)$ of 
Eq.~(\ref{Sgen2}), as a function of the parameters $(p, \gamma)$ of the thermal noise, with a 
control qubit prepared in the pure state $\ket{\psi_c}=\sqrt{p_c}\ket{0_c}+\sqrt{1-p_c}\ket{1_c}$, 
and a unitary $\mathsf{U}_\xi$ with axis $\vec{n}=\vec{e}_z$. The density operator 
$ \allowbreak 
\mathcal{S}_{00}(\rho)=\sum_{j=1}^4 \sum_{k=1}^4 
\Lambda_j \mathsf{U}_\xi \Lambda_k \mathsf{U}_\xi \rho$ $ 
\mathsf{U}_\xi^\dagger \Lambda_k^\dagger \mathsf{U}_\xi^\dagger \Lambda_j^\dagger$
can be represented as the $2\times 2$ matrix
\begin{equation}
\mathcal{S}_{00}(\rho) =
\left[ \begin{array}{lr} 
(1-\gamma)^2\rho_{00}+p(1-\gamma)\gamma+p\gamma & (1-\gamma) \rho_{01}e^{-i2\xi} \\
(1-\gamma) \rho_{01}^* e^{i2\xi}  & (1-\gamma)^2(1-\rho_{00})+(1-p)(1-\gamma)\gamma+(1-p)\gamma
\end{array} \right] \;.
\label{Apen1}
\end{equation}
The Hermitian operator $\mathcal{S}_{01}(\rho)=\sum_{j=1}^4 \sum_{k=1}^4 
\Lambda_j \mathsf{U}_\xi \Lambda_k \mathsf{U}_\xi \rho 
\mathsf{U}_\xi^\dagger \Lambda_j^\dagger \mathsf{U}_\xi^\dagger \Lambda_k^\dagger$ resulting in 
Eq.~(\ref{S01}) can be represented as the $2\times 2$ matrix
\begin{equation}
\mathcal{S}_{01}(\rho) =
\left[ \begin{array}{lr} 
b_{00}   & b_{01}\\
b_{01}^* & b_{11}
\end{array} \right] \;,
\label{Apen2b}
\end{equation}
with the matrix elements
\begin{eqnarray}
\label{Apend_b00}
b_{00} &=& 2\gamma\sqrt{1-\gamma}p(1-\rho_{00})\cos(\xi)+\bigl[1-\gamma (1-p) \bigr]^2 \rho_{00} \;,\\
\label{Apend_b01}
b_{01} &=& \bigl[(1-\gamma)e^{-i2\xi}+\gamma^2(1-p)p\bigr]\rho_{01} \;,\\
\label{Apend_b11}
b_{11} &=& 2\gamma\sqrt{1-\gamma}(1-p)\rho_{00}\cos(\xi)+(1-\gamma p)^2 (1-\rho_{00}) \;.
\end{eqnarray}
We especially obtain for Eq.~(\ref{Qfactor1}) the trace
$Q_c = \tr\bigl[ \mathcal{S}_{01}(\rho) \bigr]=b_{00}+b_{11}$ expressed in Eq.~(\ref{Qxi1}).
From these two matrices for $\mathcal{S}_{00}(\rho)$ and $\mathcal{S}_{01}(\rho)$,
the joint state $\mathcal{S}(\rho \otimes \rho_c)$ of Eq.~(\ref{Sgen2}) for the probe-control qubit
pair of the switched channel can be represented as the $4\times 4$ matrix
\begin{equation}
\mathcal{S}(\rho \otimes \rho_c) =
\left[ \begin{array}{cc} 
p_c \mathcal{S}_{00}(\rho)                & \sqrt{(1-p_c)p_c}\,\mathcal{S}_{01}(\rho)\\
\sqrt{(1-p_c)p_c}\,\mathcal{S}_{01}(\rho) & (1-p_c)\mathcal{S}_{00}(\rho)
\end{array} \right] \;.
\label{Apen4}
\end{equation}

As mentioned at the beginning of Section~\ref{meas_sec}, if only the probe qubit, that interacts
with the unitary $\mathsf{U}_\xi$, is measured, then the performance for estimating the phase $\xi$ 
is similar to a standard cascade, with standard properties. Meanwhile, if only the control qubit, 
that does not directly interact with the unitary $\mathsf{U}_\xi$, is measured, then non-standard 
performance results, with non-standard properties inaccessible to standard estimation, as analyzed 
in Section~\ref{meas_sec}. If the two qubits, probe and control, of the switched channel, are 
measured for phase estimation, the resulting performance can be analyzed by means of the joint 
state $\mathcal{S}(\rho \otimes \rho_c)$ characterized in Eq.~(\ref{Apen4}). An eigendecomposition 
of this bipartite state $\mathcal{S}(\rho \otimes \rho_c)$ would give access to its four eigenstates 
$\ket{\lambda_\ell}$ and four eigenvalues $\lambda_\ell$ determining \cite{Paris09,Chapeau15} the
quantum Fisher information 
\begin{equation}
F_q(\xi) = 2\sum_{\ell=1}^4 \sum_{m=1}^4 
\dfrac{|\braket{\lambda_\ell |\partial_\xi \mathcal{S}(\rho \otimes \rho_c) | 
\lambda_m}|^2}{\lambda_\ell + \lambda_m} 
\label{7.F1_sol2A}
\end{equation}
characterizing the performance upon measuring the qubit pair for estimating the phase $\xi$.
This, however, can hardly be accomplished analytically in general, and preclude the accessibility 
of closed-form analytical solutions as they were obtained in Section~\ref{meas_sec} for the
control qubit. Alternatively, numerical analysis can be implemented, with however a rather 
large range of configurations to be explored, according to the noise parameters $(p, \gamma)$, the 
input configuration of the qubit pair and its output measurement. A combination of the standard and 
non-standard properties can be expected to hold.
Such two-qubit approaches and their optimization remain open for further characterization.


\end{document}